\documentclass{article}%
\usepackage{amsmath}
\usepackage{amsfonts}
\usepackage{amssymb}
\usepackage{graphicx}%
\providecommand{\U}[1]{\protect\rule{.1in}{.1in}}

\begin{document}

\begin{center}
{\LARGE \ Resolving a controversy about adhesion in sliding contacts}

\bigskip\medskip

G.Cricr\`{\i}$^{1}$, M.Ciavarella$^{2,3}$

$^{1}$Universit\`{a} di Napoli Federico\ II, DII dept., P.Tecchio, 80. 80125
Napoli (It)

$^{2}$Politecnico di BARI. DMMM dept. V Orabona, 4, 70126 Bari. (It) mciava@poliba.it

$^{3}$Hamburg University of Technology, Dep Mech Eng, Am Schwarzenberg-Campus
1, 21073 Hamburg, Germany

\medskip
\end{center}

\bigskip\textbf{Abstract:} An interesting recent paper by Menga, Carbone \&
Dini (MCD, 2018, [1]), suggests that in sliding adhesive contacts, the contact
area should increase due to tangential shear stresses at the interface,
assumed to be constant and corresponding to a material constant. This is
\textit{not} observed in the known experiments, and is in sharp contrast with
\textit{all} the classical theories about the transition from stick to
sliding, both in the JKR (Griffith like) conditions which involve singular
pressure and shear, as well as in full general cohesive models. We offer a
rigorous thermodynamics calculation, which suggests in fact there is no
qualitative contrast but a very close quantitative agreement, with previous
theories. Actually, the model predicts an even stronger reduction of contact
area than predicted by Savkoor and Briggs, contrary to experimental
observations, so would certainly require some adjustements to consider
dissipative effects.

\bigskip{} \textbf{Keywords: }Adhesion, JKR model, friction, soft matter,
fracture mechanics, mixed mode, cohesive models.

\section{\bigskip{}Introduction}

A recent interesting paper by Menga, Carbone and Dini [1], applied to contact
mechanics problem in the presence of adhesion and friction, assumes no
dissipation, and yet seems to suggest that the contact area \textit{increases}
in the presence of shear tractions, in particular, assuming them constant as
apparently observed in some experiments [2], which corresponds to an effective
\textit{increase} of the mode I surface energy. In fracture mechanics, mixed
mode enhances the toughness (the generalization of the concept of surface
energy) observed in pure mode I ([3,4,5]), but this is due to dissipative
effects, and anyway we shall see it is insufficient to cause area increase (we
shall call it "MCD paradox" in the following), as indeed all the other models
of contact, either assuming singular full stick "Signorini" contact conditions
(Savkoor and Briggs, [6]) but without dissipation ("ideally brittle"), or even
more general cohesive models (Johnson, [7]) empirically adding the effect of
dissipation, invariably find \textit{reduction} of the contact area. This is
experimentally confirmed in both old [6] as well as very recent and careful
experiments ([8-13]). Dissipation, at most, \textit{limits }the contact area
reduction to a much weaker dependence on tangential force, but doesn't seem to
lead to any increase. Instead, taking into account of dissipation, the MCD
paradox would lead to even larger paradoxical increases.

We shall use here a simple energetic treatment, based on two equivalent
procedures (a more direct calculation, and another based on Legendre
transform), to offer an alternative result, which is completely in line with
existing theories and results.

\section{Models in adhesive contacts}

Let us consider a generic mode I (no shear tractions, see Fig.1) contact
problem with contact area $A$, and use the classical thermodynamic treatment
as in Maugis ([14], par.3.2) which applies to contact problem classical
fracture mechanics. The quantity $G$, which describes the variations of
elastic energy with $A$ at constant remote displacement $\delta$, is the
(elastic) energy release rate
\begin{equation}
G=\left(  \frac{\partial U_{E}}{\partial A}\right)  _{S,\delta}%
\end{equation}
where $U_{E}$ is elastic strain energy, and the derivative is calculated at
constant entropy $S$ and (remote) displacement$\ \delta$. This can be obtained
from an energy balance (Griffith balance) where the energy is the sum of
elastic strain energy $U_{E}$ and surface energy $U_{S}=-G_{Ic}A$. The
variation of the total energy at equilibrium is zero
\begin{equation}
\left(  \frac{\partial U}{\partial A}\right)  _{S,\delta}=\left(
\frac{\partial U_{E}}{\partial A}\right)  _{S,\delta}+\left(  \frac{\partial
U_{S}}{\partial A}\right)  _{S,\delta}=G-G_{Ic}=0
\end{equation}
where $G_{Ic}$ is toughness or surface energy\footnote{Under force control,
instead, we easily get an alternative%
\begin{equation}
G=\left(  \frac{\partial U_{E}}{\partial A}\right)  _{S,\delta}=\left(
\frac{\partial U_{E}}{\partial A}+\frac{\partial U_{P}}{\partial A}\right)
_{P}=G_{Ic}%
\end{equation}
}. Classical methods to compute $G$ permit to solve the problem, for example
$G$ is estimated from stress intensity factors (in the Irwin equivalent
procedure of the energetic one) if we are under Linear Elastic Fracture
mechanics framework, or with cohesive models in more general cases (as in
Johnson [7] model). The "JKR-assumptions" assume extremely short range
adhesive forces (virtually a delta-function), and hence correspond to the
classical Signorini definition of contact as either "intimate contact" or full
separation, for which the contact area is clearly defined. In cohesive models,
generally there is a continuous transition between contact and separation, the
definition of "contact area" is blurred, and the simple JKR procedure cannot
be applied. Here, JKR stands for Johnson, Kendall and Roberts [15] fundamental
contribution, which introduced Griffith-like energy balance in the world of
contact mechanics, and was later discussed to be the correct limit for soft
and large bodies, or more precisely when Tabor parameter [16] \ for the
sphere, is large
\begin{equation}
\mu_{sphere}=\left(  \frac{RG_{Ic}^{2}}{E^{\ast2}\Delta r^{3}}\right)
^{1/3}=\frac{\left(  Rl_{a}^{2}\right)  ^{1/3}}{\Delta r}=\frac{\sigma_{0}%
}{E^{\ast}}\left(  \frac{R}{l_{a}}\right)  ^{1/3}\rightarrow\infty
\label{Tabor}%
\end{equation}
where $R$ is the sphere radius, $\Delta r$ is the range of attraction of
adhesive forces, and $E^{\ast}$ the plane strain elastic modulus. $E^{\ast
}=\left(  \frac{1-\nu_{1}^{2}}{E_{1}}+\frac{1-\nu_{2}^{2}}{E_{2}}\right)
^{-1}$ and $E_{i},$ $\nu_{i}$ are the Young modulus and Poisson ratio of the
material couple. Also, $\sigma_{0}$ is the theoretical strength of the
material, and we have introduced the length $l_{a}=G_{Ic}/E^{\ast}$ as an
alternative measure of adhesion, usually a very small length scale.

\bigskip

\begin{center}
$%
\begin{array}
[c]{cc}%
{\includegraphics[
height=2.9559in,
width=5.2088in
]%
{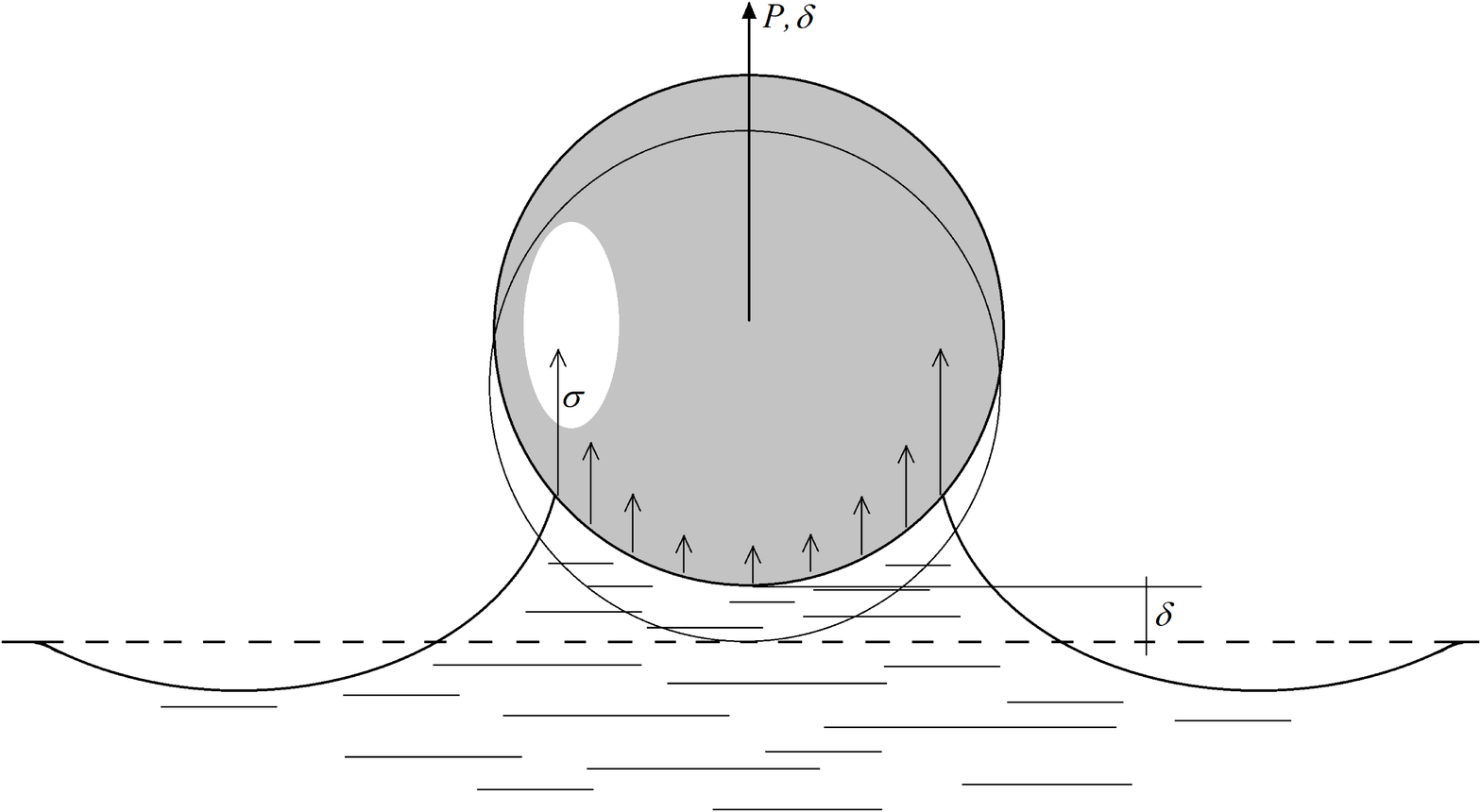}%
}
& \\
&
\end{array}
$

Fig.1. Geometry of the problem in pure mode I (without tangential forces)
\end{center}

When extending this calculation to mixed mode (Savkoor and Briggs [6]) in
terms of Irwin Stress Intensity Factors we have in principle all 3 modes of
fracture, but normally, to retain the axisymmetric simplification, we average
the values around the periphery, obtaining
\begin{equation}
G=\frac{1}{2E^{\ast}}\left[  K_{I}^{2}+\frac{2-\nu}{2\left(  1-\nu\right)
}K_{II}^{2}\right]  \label{1}%
\end{equation}
where for incompressible materials the factor $\frac{2-\nu}{2\left(
1-\nu\right)  }$ is equal to $3/2.$ Here,
\begin{equation}
K_{I}=\frac{P_{a}}{2a\sqrt{\pi a}}=\frac{p_{a}}{2}\sqrt{\pi a};\qquad
K_{II}=\frac{T}{2a\sqrt{\pi a}}=\frac{\tau_{m}}{2}\sqrt{\pi a} \label{Ki}%
\end{equation}
where $P=P_{H}-P_{a}$ while $P_{H}=\frac{4E^{\ast}a^{3}}{3R}$ is a compressive
Hertzian load and $P_{a}$ is responsible of the contact edge singularity.
Here, $a$ is the radius of the circular contact area, $a=\sqrt{A/\pi}$. Hence,
this "ideally brittle fracture" model equilibrium dictates
\begin{equation}
G_{c}=G_{Ic}=\frac{1}{2E^{\ast}}\left[  \left(  \frac{p_{a}}{2}\right)
^{2}+\frac{3}{2}\left(  \frac{\tau_{m}}{2}\right)  ^{2}\right]  \pi a
\end{equation}
and $\ $therefore the contact area will follow the JKR equation, but with a
reduced effective surface energy
\begin{equation}
G_{Ic,eff}=G_{Ic}-\frac{3\pi}{8}\frac{\tau_{m}^{2}a}{E^{\ast}} \label{Savkoor}%
\end{equation}
which shows also a size effect. Experiments of Savkoor and Briggs clearly
evidenced a \textit{reduction of the contact area} when tangential load was
applied, but less than expected from the model, and this clearly indicated
dissipative processes.

In the interesting MCD paradox paper ([1]), their calculation leads to their
eqt.26 (both under force, or under displacement control in mode I)
\begin{equation}
G_{Ic,eff}=G_{Ic}+\frac{4\tau^{2}a}{\pi E^{\ast}}\quad\label{menga}%
\end{equation}
where $\tau$ is here a material constant shear stress in sliding, i.e. the
effective surface energy (or toughness of the interface) is \textit{increased}%
, rather than decreased in Savkoor's theory (\ref{Savkoor}), and curiously of
a very similar quantity, as $\frac{3\pi}{8}=\allowbreak1.\,\allowbreak178\,$
while $\frac{4}{\pi}=\allowbreak1.\,\allowbreak273\,$, the difference perhaps
being that MCD doesn't take into account of mode III and of averaging around
the periphery. In MCD, it appears also surprising in that there would be an
"effective adhesion", as $G_{Ic,eff}\rightarrow\frac{4\tau_{0}^{2}a}{\pi
E^{\ast}}$, even in the absence of adhesion: who provides this energy?\ It
seems created out of nowhere. Contrast with \textit{all} experiments we know
is also striking [6-13]. We shall therefore proceed to a rigorous
thermodynamic treatment from first principles.

\bigskip

\section{A first energy calculation}

In order to obtain the case of constant shear stress of the MCD paradox, as a
first method, we can impose a direct minimization on the total energy $U$
under variations of the relevant extensive parameters. Consider $U\left(
S,\delta,\tau,A\right)  =U_{E}\left(  S,\delta,\tau,A\right)  +U_{S}\left(
A\right)  $ in general as a function of entropy $S$, a contact indentation
displacement $\delta$, a shear stress $\tau$, and the contact area $A$. We
shall neglect variations of entropy, considering purely reversible
transformations, as the MCD paradox does. Further, we shall distinguish the
terms due to pressures and shear stresses with superscript $"N"$ and $"T"$, in
other words%
\begin{equation}
U_{E}=U_{E}^{N}+U_{E}^{T}%
\end{equation}

Notice that the two energy contributions are completely separated because the
contact is assumed to be that of a rigid sphere against an elastomer, which
has Poisson's ratio $v=0.5$. Indeed, more general conditions could be assumed
when Dundurs' second constant is equal to zero (Barber, [17]).

The total differential of energy is then%
\begin{equation}
dU=\left(  \frac{\partial U_{E}^{N}}{\partial\delta}\right)  _{A,\tau}%
d\delta+\left(  \frac{\partial\left[  U_{E}^{N}+U_{E}^{T}+U_{S}\right]
}{\partial A}\right)  _{\delta,\tau}dA+\left(  \frac{\partial U^{T}}%
{\partial\tau}\right)  _{A,\delta}d\tau\label{U1}%
\end{equation}
where as standard\ notation, $\left(  {}\right)  _{\delta,\tau}$ indicates
that the derivative should be computed with constant $\delta,\tau$. Also,
$U^{T}=U_{E}^{T}=\frac{1}{2}\tau W$, as the strain energy is easily computed
in this case from the work of the shear tractions in the contact area and the
tangential displacements, where $W$ is displaced volume $W=Aw$, and $w$ is the
mean horizontal displacement which we can write as $w=k_{w}\tau A^{1/2}%
/E^{\ast}$, where $k_{w}$ is a dimensionless coefficient of order 1, whose
exact value is unimportant here (MCD report $k_{w}=\frac{8}{\left(
\pi\right)  ^{3/2}}\simeq1.\,\allowbreak437$). Then, we immediately find
\begin{align}
\left(  \frac{\partial U_{E}^{N}}{\partial\delta}\right)  _{A,\tau}  &  =P\\
\left(  \frac{\partial U^{T}}{\partial\tau}\right)  _{A,\delta}  &  =\left(
\frac{\partial U_{E}^{T}}{\partial\tau}\right)  _{A,\delta}=W=Aw\\
\left(  \frac{\partial U_{S}}{\partial A}\right)  _{\delta,\tau}  &
=-G_{Ic}\\
\left(  \frac{\partial U_{E}^{T}}{\partial A}\right)  _{\delta,\tau}  &
=\left(  \frac{\partial\left[  \frac{1}{2}\tau^{2}A^{3/2}k_{w}/E^{\ast
}\right]  }{\partial A}\right)  _{\delta,\tau}=\frac{3}{4}\tau w
\end{align}

Hence, substituting back in (\ref{U1})
\begin{equation}
dU=Pd\delta+\left[  \left(  \frac{\partial U_{E}^{N}}{\partial A}\right)
_{\delta,\tau}+\left(  \frac{\partial U_{E}^{T}}{\partial A}\right)
_{\delta,\tau}-G_{Ic}\right]  dA+Wd\tau
\end{equation}
but as we assumed $\tau=const$, $\delta=const$ then%
\begin{equation}
\left(  dU\right)  _{\delta,\tau}=\left[  \left(  \frac{\partial U_{E}^{N}%
}{\partial A}\right)  _{\delta,\tau}+\left(  \frac{\partial U_{E}^{T}%
}{\partial A}\right)  _{\delta,\tau}-G_{Ic}\right]  dA
\end{equation}
the energy is minimum for $dU=0$ or
\begin{equation}
G_{I}=\left(  \frac{\partial U_{E}^{N}}{\partial A}\right)  _{\delta,\tau
}=G_{Ic}-\left(  \frac{\partial U_{E}^{T}}{\partial A}\right)  _{\delta,\tau
}=G_{Ic}-\frac{3\sqrt{\pi}k_{w}}{4}\frac{\tau^{2}a}{E^{\ast}}=G_{Ic,eff}%
\end{equation}
and therefore we have the problem is equivalent to a mode I problem with
\textit{reduced effective surface energy}. Notice incidentally that this
generalizes trivially the definition of energy release rate. The result
exactly corresponds to Savkoor and Briggs [6] when $\tau=\tau_{m}$ (i.e. in
the limit of full sliding), and for $k_{w}=\frac{3\pi}{8}\frac{4}{3\sqrt{\pi}%
}=\allowbreak0.89,$ whereas we have reported $k_{w}=\frac{8}{\left(
\pi\right)  ^{3/2}}\simeq1.\,\allowbreak437$, which means the models are even
\textit{quantitatively} very close, but actually the present model would lead
to an even stronger reduction of contact area than predicted by Savkoor and
Briggs [6], contrary to experimental observations. The result of this more
direct energy calculation are also qualitatively close to a full cohesive
(more general) model of Johnson [7] in the limit of full sliding and when
dissipation is neglected\footnote{Johnson suggests Savkoor and Briggs case is
when his function $f_{1}=1$ or $\alpha=1+1/2g$ (where $g=\frac{\tau
_{0}\overline{s}_{0}}{\sigma_{0}h_{0}}$, being $\tau_{0},\sigma_{0}$ cohesive
stresses, and $\overline{s}_{0},h_{0}$ are the limit displacements of the
Maugis-Dugdale equivalent model in shear and opening, respectively). See the
original paper for details.}, as Johnson's results clearly show not much
difference from the Savkoor Briggs results even when we are quite far from the
assumptions of the Griffith-LEFM-JKR case both for mode II and/or for mode I.

The present model holds up to $G_{Ic,eff}>0$. Hence, when
\begin{equation}
a>a_{0}=\frac{4}{3\sqrt{\pi}k_{w}}\frac{E^{\ast}G_{Ic}}{\tau^{2}}=\frac
{2}{3\sqrt{\pi}k_{w}}\left(  \frac{K_{Ic}}{\tau}\right)  ^{2}%
\end{equation}
one should rather expect that adhesion is completely destroyed, leading to the
standard Hertzian solution. This is quite easy for stiff materials. For
example, in the experiments in UHV in the AFM by Carpick \textit{et al.} [18],
it was measured $G_{Ic}=0.19J/m^{2}$ and a uniform frictional stress
$\tau=0.84GPa$. Also, $E^{\ast}=44GPa$ for a platinum tip contact with mica,
and hence $a_{0}\sim10nm$. In very soft materials, we can reach much higher
values of contact area with the presence of friction and adhesion.

\section{\bigskip Legendre transform alternative procedure}

We can build a thermodynamic potential formulation (see for example Maugis
[14], par.3.2) where explicit variables of the problem are $\left(
\delta,w,A\right)  $ where $\delta$ is the vertical indentation, $w$ is the
tangential imposed displacement, and $A$ is contact area, by considering first
the internal energy $\phi$
\begin{equation}
\phi=U=U_{E}\left(  \delta,w,A\right)  +U_{S}\left(  A\right)  =U_{E}%
^{N}\left(  \delta,A\right)  +U_{E}^{T}\left(  w,A\right)  -G_{Ic}A
\end{equation}
where notice that we have continued to explicitly split the normal and
tangential components of energy $U_{E}^{N}\left(  \delta,A\right)  ,U_{E}%
^{T}\left(  w,A\right)  $ since the elastic contact problem is uncoupled.
However, since we want to switch the control on $\left(  \delta,\tau,A\right)
$ rather than $\left(  \delta,w,A\right)  $, we minimize a new thermodynamic
potential obtained by considering the Legendre transform
\begin{align}
\psi &  =\phi^{\ast}=U-\delta\left.  \frac{\partial U_{E}}{\partial\delta
}\right\vert _{\left(  \tau,A\right)  }^{eq}-\tau\left.  \frac{\partial
U_{E}^{T}}{\partial\tau}\right\vert _{\left(  \delta,A\right)  }%
^{eq}\nonumber\\
&  =U_{E}+U_{E}^{T}-G_{Ic}A-P^{eq}\delta-W^{eq}\tau
\end{align}
where we wrote the superscript "eq" to remind that the partial derivatives are
computed at equilibrium. This thermodynamic potential represents the energy
which can effectively when we impose the contraints $d\delta=d\tau=0$. Hence,
writing the minimum conditions on the thermodynamic potential represent the
equilibrium conditions
\begin{align}
\left.  \frac{\partial\psi}{\partial\delta}\right\vert _{\left(
A,\tau\right)  }  &  =\left.  \frac{\partial U_{E}^{N}}{\partial\delta
}\right\vert _{\left(  A,\tau\right)  }-P^{eq}=0\\
\left.  \frac{\partial\psi}{\partial\tau}\right\vert _{\left(  A,\delta
\right)  }  &  =\left(  \frac{\partial U_{E}^{T}}{\partial\tau}\right)
_{A,\delta}-W^{eq}=0
\end{align}
and finally
\begin{align}
\left.  \frac{\partial\psi}{\partial A}\right\vert _{\left(  \delta
,\tau\right)  }  &  =\left.  \frac{\partial U_{E}^{N}}{\partial A}\right\vert
_{\left(  \delta,\tau\right)  }+\left.  \frac{\partial U_{E}^{T}}{\partial
A}\right\vert _{\left(  \delta,\tau\right)  }-G_{Ic}-P\left.  \frac
{\partial\delta}{\partial A}\right\vert _{\left(  \delta,\tau\right)
}-W\left.  \frac{\partial\tau}{\partial A}\right\vert _{\left(  \delta
,\tau\right)  }\\
&  =\left(  \frac{\partial U_{E}^{N}}{\partial A}\right)  _{\delta,\tau
}-\left(  G_{Ic}-\left(  \frac{\partial U_{E}^{T}}{\partial A}\right)
_{\delta,\tau}\right)
\end{align}
which corresponds to the same result as the previous direct calculation,
obviously. The difference with respect to the other procedure is that with the
Legendre transform, we have defined in a single thermodynamic potential a
single function "free energy" to minimize, which contains all the equilibrium
conditions of the system, given the other constraints .

\section{Conclusion}

The thermodynamics simple treatment for JKR type of adhesion, with a constant
value of shear stress in the contact area, has shown results that are
qualitatively and quantitatively similar to existing to existing Savkoor and
Briggs or Johnson cohesive models, namely that there is a large
\textit{reduction} of the contact area under reversible "ideally brittle"
conditions with no dissipative effects, as in the context of the MCD paradox
theory. Even more than this, the model shows an even larger reduction of
contact area than predicted by Savkoor and Briggs, and to fit experimental
observations, some adjustements are needed to consider dissipative effects.
This resolves a controversy generated in the interesting MCD paper, who had
suggested instead that the contact area should increase, a rather surprising
result considering it was in contrast with all previous theories, and all
previous experiments.

\section{\bigskip References}

[1] Menga, N., Carbone, G., \& Dini, D. (2018). Do uniform tangential
interfacial stresses enhance adhesion?. Journal of the Mechanics and Physics
of Solids, 112, 145-156.

\bigskip

[2] Chateauminois, A., \& Fretigny, C. (2008). Local friction at a sliding
interface between an elastomer and a rigid spherical probe. The European
Physical Journal E: Soft Matter and Biological Physics, 27(2), 221-227.

[3] Evans, A. G., \& Hutchinson, J. W. (1989). Effects of non-planarity on the
mixed mode fracture resistance of bimaterial interfaces. Acta Metallurgica,
37(3), 909-916.

[4] Cao, H. C., \& Evans, A. G. (1989). An experimental study of the fracture
resistance of bimaterial interfaces. Mechanics of materials, 7(4), 295-304.

\bigskip

[5] Hutchinson, J. W. (1990). Mixed mode fracture mechanics of interfaces.
Metal-ceramic interfaces, 295-306.

\bigskip

[6] Savkoor, A. R. \& Briggs, G. A. D. (1977), The effect of a tangential
force on the contact of elastic solids in adhesion. Proc. R. Soc. Lond. A 356, 103--114.

\bigskip\lbrack7] Johnson, K. L., (1997), Adhesion and friction between a
smooth elastic spherical asperity and a plane surface. In Proceedings of the
Royal Society of London A453, No. 1956, pp. 163-179).

\bigskip

[8] Sahli, R. Pallares, G. , Ducottet, C., Ben Ali, I. E. , Akhrass, S. Al ,
Guibert, M. , Scheibert J. , (2018) Evolution of real contact area under
shear, Proceedings of the National Academy of Sciences, 115 (3) 471-476; DOI: 10.1073/pnas.1706434115

[9] Ciavarella, M. (2018). Fracture mechanics simple calculations to explain
small reduction of the real contact area under shear. Facta universitatis,
series: mechanical engineering, 16(1), 87-91.

[10] Papangelo, A., \& Ciavarella, M. (2019). On mixed-mode fracture mechanics
models for contact area reduction under shear load in soft materials. Journal
of the Mechanics and Physics of Solids, 124, 159-171.

[11] Waters JF, Guduru PR, (2009), Mode-mixity-dependent adhesive contact of a
sphere on a plane surface. Proc R Soc A 466:1303--1325.

[12] Sahli, R., Pallares, G., Papangelo, A., Ciavarella, M., Ducottet, C.,
Ponthus, N., \& Scheibert, J. (2019). Shear-induced anisotropy in rough
elastomer contact. Physical Review Letters, 122(21), 214301.

[13] Papangelo, A., Scheibert, J., Sahli, R., Pallares, G., \& Ciavarella, M.
(2019). Shear-induced contact area anisotropy explained by a fracture
mechanics model. Physical Review E, 99(5), 053005.

[14] Maugis, D. (2013). Contact, adhesion and rupture of elastic solids (Vol.
130). Springer Science \& Business Media.

[15] Johnson, K. L., Kendall, K. \& Roberts, A. D. (1971) Surface energy and
the contact of elastic solids. Proc. R. Soc. Lond. A 324, 301--313.

[16] Tabor, D. (1977) Surface forces and surface interactions. J. Colloid
Interface Sci. 58, 2.

[17] Barber, JR. (2018). Contact Mechanics. Springer, New York.

[18] Carpick, R. W., Agrait, N., Ogletree, D. F. \& Salmeron, M. (1996)
Variation of the interfacial shear strength and adhesion of a nanometer sized
contact. Langmuir 12, 3334--3340.

\end{document}